# Intrinsic Exchange Bias from Interfacial Reconstruction in an Epitaxial $Ni_xCo_yFe_{3-x-y}O_4(111)/\alpha$-$Al_2O_3(0001)$ Thin Film Family

**Detian Yang[1,2], Yaohua Liu[3] and Xiaoshan Xu[2,4,*]**

[1] Shanghai Key Laboratory of High Temperature Superconductors, Department of Physics, Shanghai University, Shanghai 200444, China
[2] Department of Physics and Astronomy, University of Nebraska, Lincoln, Nebraska 68588, USA
[3] Second Target Station, Oak Ridge National Laboratory, Oak Ridge, Tennessee 37830, USA
[4] Nebraska Center for Materials and Nanoscience, University of Nebraska, Lincoln, Nebraska 68588, USA

E-mail: xiaoshan.xu@unl.edu

## Abstract

Intrinsic exchange bias is known as the unidirectional exchange anisotropy that emerges in a nominally single-component ferro-(ferri-)magnetic system. In this work, with magnetic and structural characterizations, we demonstrate that intrinsic exchange bias is a general phenomenon in (Ni, Co, Fe)-based spinel oxide films deposited on $\alpha$-$Al_2O_3(0001)$ substrates, due to the emergence of a rock-salt interfacial layer consisting of antiferromagnetic CoO from interfacial reconstruction. We show that in $Ni_xCo_yFe_{3-x-y}O_4(111)/\alpha$-$Al_2O_3(0001)$ films, intrinsic exchange bias and interfacial reconstruction have consistent dependences on Co concentration $y$, while the Ni and Fe concentration appears to be less important. This work establishes a family of intrinsic exchange bias materials with great tunability by stoichiometry and highlights the strategy of interface engineering in controlling material functionalities.

## 1. Introduction

At an interface between two oxides of distinctive structural, charge, spin or orbital orders, dramatic reconstruction[1] can be accommodated and induce novel structural, electronic, and magnetic states. Examples include interfacial MnO double layers with charge polarization in $YMnO_3/Al_2O_3$ [2], quasi-2D electron gas between insulating $LaAlO_3$ and $SrTiO_3$[3], interfacial superconductivity between insulating $La_2CuO_4$ and metallic $(La, Sr)_2CuO_4$[4], and ferromagnetic metal state with colossal magnetoresistance in the superlattice of antiferromagnetic insulators $LaMnO_3$ and $SrMnO_3$[5].

Exchange bias[6,7] is a unidirectional anisotropy generated by interfacial exchange interaction between a ferromagnet and an antiferromagnet and works as the essential principle of stabilizing soft ferromagnetic components in magnetic read heads and pinning harder reference layers in spin valve devices. Intrinsic exchange bias[8] is the phenomenon that exchange bias arises in a nominally single-component ferro/ferri-magnetic material grown on a non-magnetic substrate. Without intentional fabrication of antiferromagnetic components, intrinsic exchange bias phenomena can simplify device designs in magnetic storage and spintronics in view of ever-increasing demands for miniaturization and cost-effect device manufacturing. To date, intrinsic exchange bias has been reported in various heterostructures such as $LaNiO_3/LaMnO_3$ superlattices[9], $La_{2/3}Sr_{1/3}MnO_3/LaSrAlO_4$ [10], $SrRuO_3/LaAlO_3$[11], Fe/MgO[12] and $La_{0.67}Sr_{0.33}MnO_{3-\delta}/SrTiO_3$[13]. However, most intrinsic exchange bias studies in thin films remain sporadic and provide poor tunability. And none of these reported materials' exchange biases surpass their coercivities, failing to meet the necessary criterion for application.

Transition metal spinel $TM_3O_4$ (TM=Fe, Co, Ni) such as $Fe_3O_4$, $CoFe_2O_4$ and $NiCo_2O_4$ have been intensively studied for applications in chemical/biosensors[14], energy storage[15], electromechanical devices[16] and spintronic devices[17,18], owning to the broad scope of their flexible magnetic, electronic, optical and chemical properties. Despite the large structural difference between cubic $TM_3O_4$ and rhombohedral $\alpha$-$Al_2O_3$, $TM_3O_4$ thin films such as $Fe_3O_4$[19], $CoFe_2O_4$[20] and $NiCo_2O_4$[21] can be epitaxially grown on $\alpha$-$Al_2O_3$. We have previously reported that, in $CoFe_2O_4(111)/\alpha$-$Al_2O_3(0001)$ and $NiCo_2O_4(111)/\alpha$-$Al_2O_3(0001)$ thin films, an interfacial layer containing the antiferromagnetic CoO develops from interfacial structural reconstruction, where a colossal intrinsic exchange bias as large as 7 kOe and an exchange bias beyond the coercivity are observed, respectively[8]. Here we demonstrate that such intrinsic exchange bias emerges in spinel thin films $Ni_xCo_yFe_{3-x-y}O_4(111)/\alpha$-$Al_2O_3(0001)$ ($0 \leq x+y \leq 3$) from interfacial reconstruction as the Co concentration y lies in the range $0.15 \leq y \leq 2$; and are greatly tunable by Co concentration.

## 2. Experimental Methods

(111)-oriented $NiFe_2O_4$, $Ni_{0.95}Co_{0.15}Fe_{1.95}O_4$, $Ni_{0.75}Co_{0.75}Fe_{1.5}O_4$, $Ni_{0.67}CoFe_{1.33}O_4$, $CoFe_2O_4$, $Ni_{0.5}Co_{1.5}FeO_4$, $NiCo_2O_4$, $Ni_{0.15}Co_{2.55}Fe_{0.3}O_4$ thin films were grown on $\alpha$-$Al_2O_3$ (0001) substrates by pulsed laser deposition (PLD) with an oxygen pressure of 5 mTorr at 520°C, 440°C, 390°C, 340°C, 530°C, 330°C, 300°C and 20°C, respectively. All targets were fabricated by solid state reaction from $Fe_2O_3$, NiO and $Co_2O_3$ powders with a purity of higher than 99.9%. The KrF excimer laser of wavelength 248 nm was employed to ablate the targets with a pulse energy of 120±10mJ and a repetition rate of 4 Hz. Growth temperatures for all materials have been optimized. The growth processes were in-situ monitored by a reflection high energy electron diffraction (RHEED) system. For every sample, we in-situ recorded the RHEED images along $Ni_xCo_yFe_{3-x-y}O_4[\bar{1}\bar{1}2]$ / $\alpha$-$Al_2O_3[1\bar{1}00]$ every second.

The out-of-plane x-ray diffraction (XRD) and x-ray reflectivity (XRR) were conducted by a Rigaku SmartLab x-ray diffractometer (copper K-α source, X-ray wavelength 1.5406 Å); the film thicknesses were extracted from the XRR data. The in-plane crystal structure was studied by analyzing thickness-resolved RHEED patterns.

The magnetic hysteresis loops were measured from 5K to 300K in a superconducting quantum interfere device (SQUID) system after being cooled down from 320 K in ±4 T and ±7 T fields. Exchange bias $H_E$ and coercivity $H_C$ are defined conventionally by $|H_{01}+H_{02}|/2$ and $|H_{01}-H_{02}|/2$, respectively, where $H_{01}$ and $H_{02}$ are the two field values at which magnetization $M(H)$=0.

## 3. Universality of intrinsic exchange bias in spinel $Ni_xCo_yFe_{3-x-y}O_4$ films

The ideal spinel structure of $TM_3O_4$ features a face-centred-cubic oxygen ion sublattice with half of its octahedral sites and one eighth of its tetrahedral sites occupied by smaller TM cations[22]. TM cations arrange in such a way that the lattice constant of TM cation sublattice is the twice as much as that of the oxygen ion sublattice. In this work, eight members of $Ni_xCo_yFe_{3-x-y}O_4$ (111) ($0<x+y\leq3$) thin films, i.e., $NiFe_2O_4$, $Ni_{0.95}Co_{0.15}Fe_{1.95}O_4$, $Ni_{0.75}Co_{0.75}Fe_{1.5}O_4$, $Ni_{0.67}CoFe_{1.33}O_4$, $CoFe_2O_4$, $Ni_{0.5}Co_{1.5}FeO_4$, $NiCo_2O_4$, $Ni_{0.15}Co_{2.55}Fe_{0.3}O_4$, were epitaxially grown on $\alpha$-$Al_2O_3$(0001) substrates by pulsed laser deposition and the growth processes were monitored in situ by RHEED. The cation ratios were chosen so that samples have varying concentrations of Fe, Co and Ni; especially the Co ratio covers a large range from 0 to 2.55. The films' spinel structure is confirmed by the θ-2θ scan of XRD of $Ni_xCo_yFe_{3-x-y}O_4$($0\leq x+y\leq3$) films in Fig. 1(a) and the typical RHEED patterns along two perpendicular in-plane directions shown in Fig. 1(b) (see Fig. S1 in Supplemental Materials for RHEED patterns of all eight $Ni_xCo_yFe_{3-x-y}O_4$ films). In *x-y* phase diagram Fig. 2(a), the emergence of the intrinsic exchange bias of $Ni_xCo_yFe_{3-x-y}O_4$ films is demonstrated, where the red balls denote the film members with exchange bias and the blue balls indicate the members of which no exchange bias were observed. Shown in Fig. 2(b)-(e) are typical in-plane hysteresis loops of 4 materials measured at 20K after the samples were cooled down in $\pm70$ kOe fields (for hysteresis loops of more materials, see Fig. S2 in Supplemental Materials). Clearly, except $NiFe_2O_4$ and $Ni_{0.15}Co_{2.55}Fe_{0.3}O_4$, all other samples in $Ni_xCo_yFe_{3-x-y}O_4$ series show significant exchange bias. $NiCo_2O_4$ stands out as its $H_E$ surpasses $H_C$ as shown in Fig. 2(e) [23] . It is worth mentioning that, as shown in Fig.S3 and discussed in Section 3 in the Supplementary Materials, the relatively thin samples (about 10 nm) of some films such as $Ni_{0.75}Co_{0.75}Fe_{1.5}O_4$ and $Ni_{0.67}CoFe_{1.33}O_4$ demonstrate intriguing large nominal exchange bias of around 10 kOe as a result of the significant vertical shifts of hysteresis loops along the magnetization axis. The inset of Fig. 2(a) illustrates the proposed realistic structure of $Ni_xCo_yFe_{3-x-y}O_4(111)/\alpha$-$Al_2O_3(0001)$ thin films with an interfacial layer from reconstruction. As already demonstrated in [8,23], this interfacial layer is proposed to contain antiferromagnetic CoO which pins the magnetization of the ferrimagnetic $Ni_xCo_yFe_{3-x-y}O_4$ layer to produce the intrinsic exchange bias.

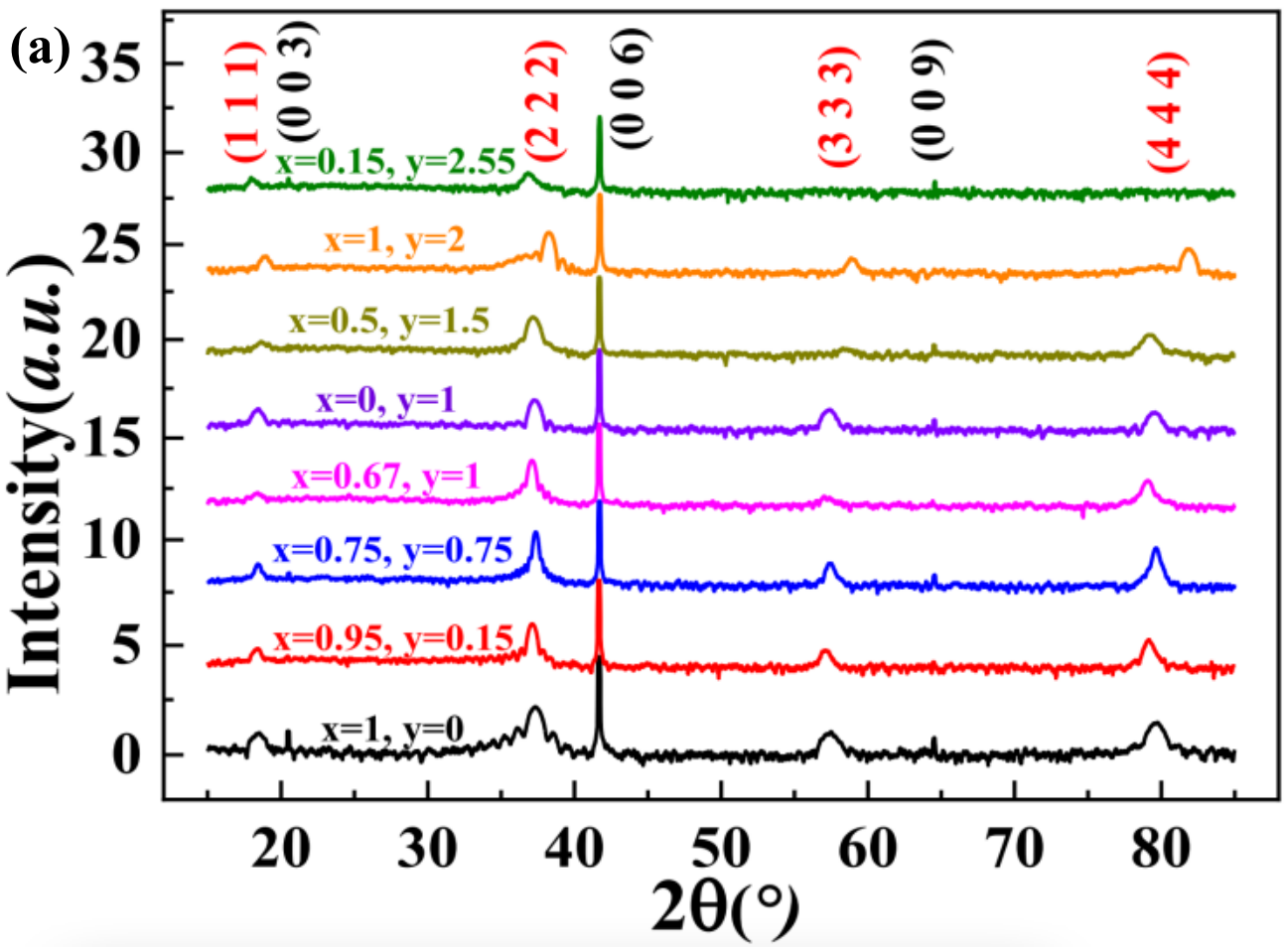

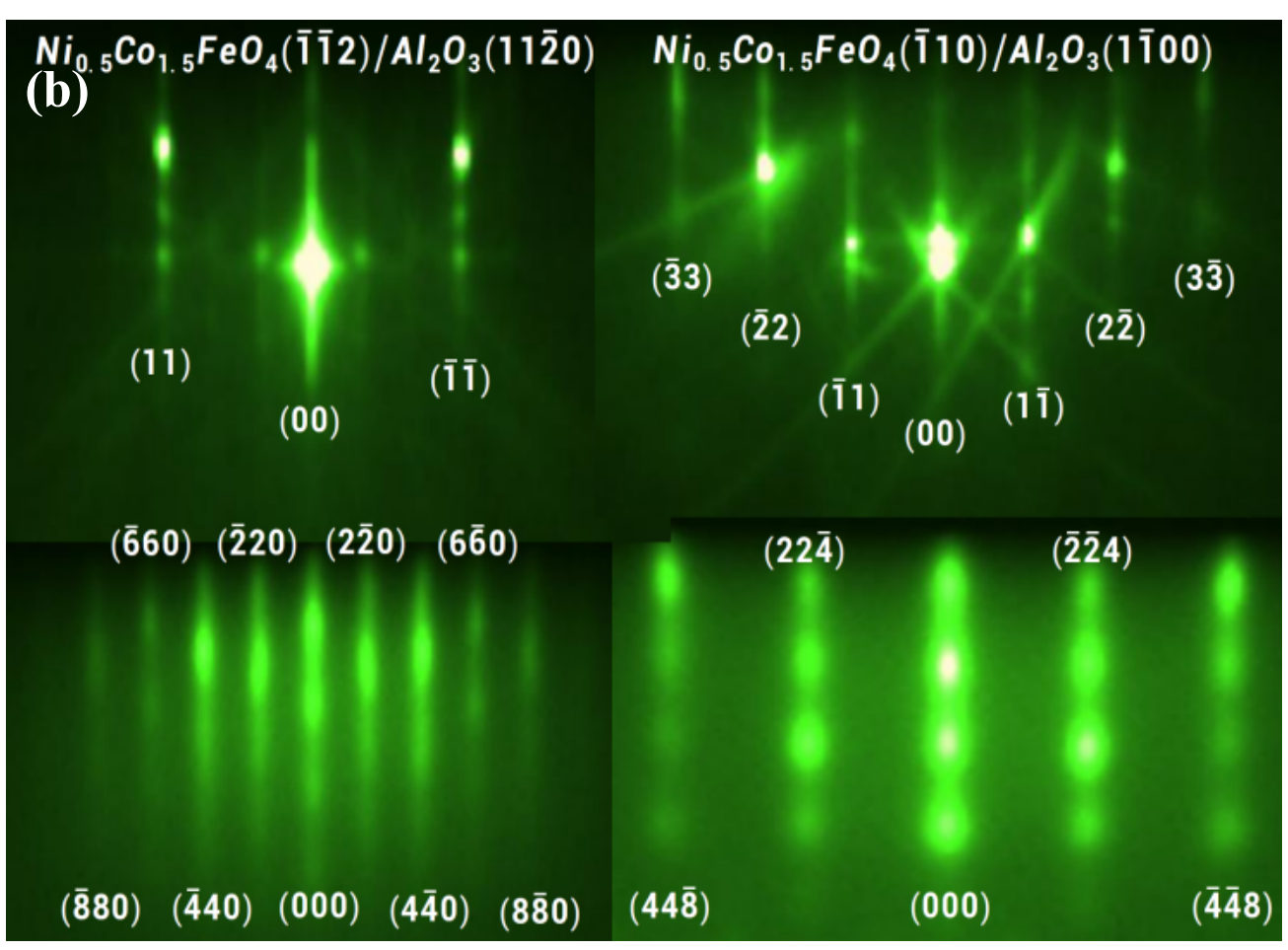

FIG.1. **(a)** θ-2θ XRD of $Ni_xCo_yFe_{3-x-y}O_4$($0 \leq x+y \leq 3$) films. The thicknesses of the samples from bottom to top are 11 nm, 18nm, 32 nm, 21nm, 13nm,18nm, 15nm and, 18nm, respectively. **(b)** RHEED patterns of $\alpha$-$Al_2O_3$ and a 18nm $Ni_{0.5}Co_{1.5}FeO_4$ film along two perpendicular in-plane directions $NiCo_2O_4$ $[\bar{1}10]$ and $[\bar{1}\bar{1}2]$ of the sample in (a).

## 4. Interfacial nature and blocking temperature of intrinsic exchange bias

The interfacial nature of intrinsic exchange biases in $Ni_xCo_yFe_{3-x-y}O_4$ thin films is indicated by the thickness-dependence of $H_E$ of $NiCo_2O_4$, $CoFe_2O_4$ and $Ni_{0.75}Co_{0.75}Fe_{1.5}O_4$ in Fig. 3(a), where $t_{FM}$, $t_I$, and $t$, are the thickness of the ferrimagnetic part of the film, the interfacial layer, and the whole film, respectively. Here, $CoFe_2O_4$ samples were grown in 10mTorr $O_2$, while $NiCo_2O_4$ and $Ni_{0.75}Co_{0.75}Fe_{1.5}O_4$ samples were grown in 5mTorr $O_2$. Evidently, as the thickness increases, $H_E$ of all three materials decrease steadily here. $H_E(t_{FM})$ of all cases can be fitted into the power law relations with a power of 0.9, 0.4 and 0.5 for $NiCo_2O_4$, $CoFe_2O_4$ and $Ni_{0.75}Co_{0.75}Fe_{1.5}O_4$, respectively. The deviation from the power 1 in the traditional Meiklejohn-Bean model [6] could be due to the smearing interface between the interfacial layer and the ferrimagnetic layer, which will be shown in Fig.4.

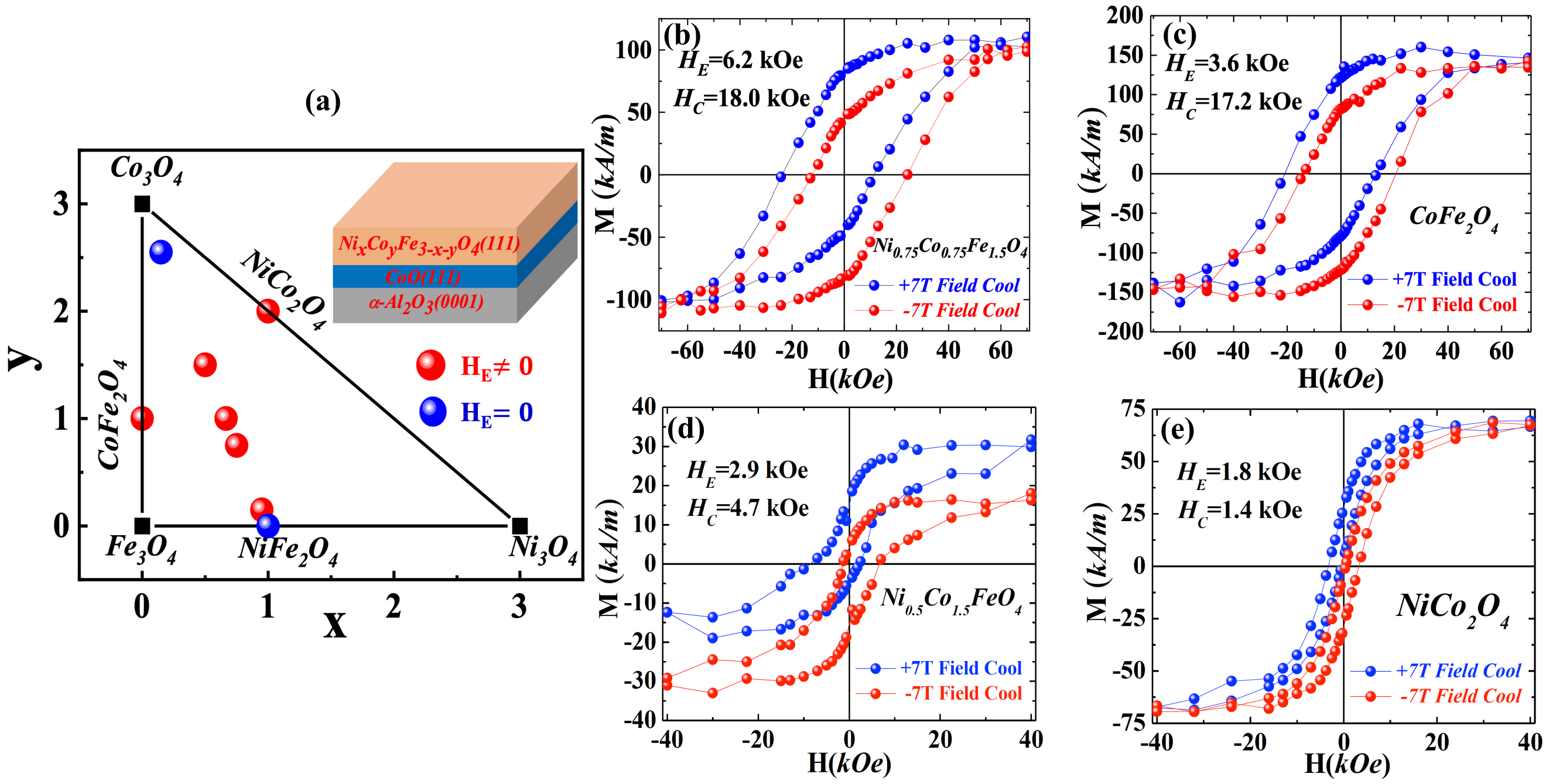

FIG. 2. **(a)** Emergence of Intrinsic exchange bias in spinel $Ni_xCo_yFe_{3-x-y}O_4$ ($0\leq x+y\leq3$) thin films shown in the *x-y* phase diagram. The films in which nonzero exchange bias ($H_E\neq 0$) was observed are denoted by red solid circles, while those do not exhibit exchange bias ($H_E= 0$) by blue solid circles. Experimentally, any exchange bias value smaller than 0.2 kOe was taken as a result of experimental errors. The inset represents the proposed structure of $Ni_xCo_yFe_{3-x-y}O_4$ thin films with an interfacial layer shown with the blue colour. **(b)-(e)** Featured magnetic in-plane hysteresis loops of $Ni_xCo_yFe_{3-x-y}O_4$ films (grown in 5 mTorr oxygen gas) measured at 20K after cooling in ±7T fields: **(b)**16nm $Ni_{0.75}Co_{0.75}Fe_{1.5}O_4$; **(c)** 9 nm $CoFe_2O_4$; **(d)**11nm $Ni_{0.5}Co_{1.5}FeO_4$; **(e)** 10nm $NiCo_2O_4$. $H_E$ and $H_C$ are exchange bias and coercivity, respectively.

Temperature-dependent exchange bias and coercivity of a 10 nm $NiCo_2O_4$ film, a 10 nm $CoFe_2O_4$ film and a 12 nm $Ni_{0.75}Co_{0.75}Fe_{1.5}O_4$ film in Fig. 3(b) demonstrate that both $H_E$ and $H_C$ in general increase with the decreasing of the temperature except the abrupt drops in $H_E$ and $H_C$ for $T$<20K in $NiCo_2O_4$ and $CoFe_2O_4$ films coinciding with the sharp jump of their saturation magnetizations for $T$<20K in Fig. 3(c). Presumably, this phenomenon could be a magnetic phase transition from the low-spin state to the high spin state. It might be induced by the strain effect because it was not observed in thick samples. Note that such dramatic changes of $H_E$, $H_C$ and $M_S$ for $T$<20K were not observed in the case of $Ni_{0.75}Co_{0.75}Fe_{1.5}O_4$ (Fig.3(b) & (c)), which implies that such a transition is closely related to the stoichiometry of the films. Since thermal energy can disrupt magnetic order and so weaken the interfacial exchange coupling, exchange bias normally can only survive below certain temperature, which is known as the blocking temperature of exchange bias. From Fig.3(b), clearly, the blocking temperatures of $H_E$ for all three materials are consistently around 300 K, suggesting a similar or same exchange bias mechanism. Since according to Fig.3(c), room-temperature magnetizations of all three kinds of thin films stay still quite large, the Curie temperatures of $NiCo_2O_4$, $CoFe_2O_4$ and $Ni_{0.75}Co_{0.75}Fe_{1.5}O_4$ are expected to go way beyond 300K. Therefore, the blocking temperatures here mainly reflect the degradation of the interfacial layer's antiferromagnetic order, and so, are supposed to be close to the Néel temperatures of the presumptive antiferromagnetic interfacial layer. Common antiferromagnetic oxides of Fe, Co and Ni include $\alpha$-$Fe_2O_3$ ($T_N$~950K)[24], FeO ($T_N$~190K)[25], CoO ($T_N$~290K)[26], $Co_3O_4$ ($T_N$~40K)[27] and NiO ($T_N$~90K)[28]. As only the Néel temperature of CoO matches the vanishing temperature of $H_E$ in the Fig.3(b), this supports the structural model in the inset of Fig. 2(a) with an interfacial layer dominated by CoO.

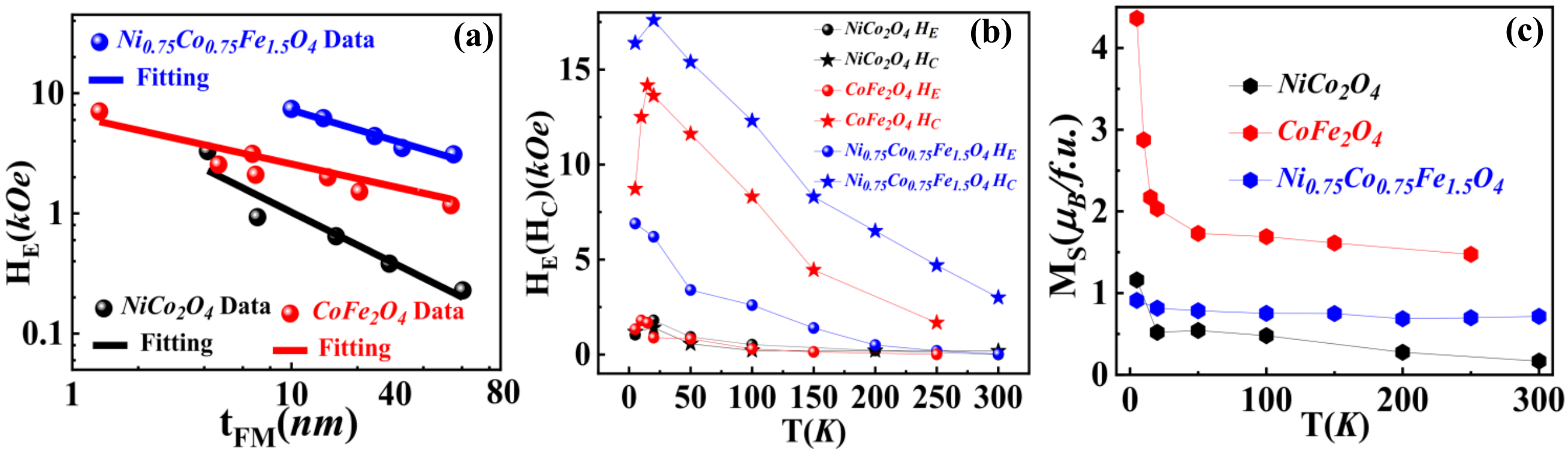

FIG. 3. **(a)** Thickness-dependent exchange bias $H_E$ (circles) and their fittings (lines) of $NiCo_2O_4$ (black), $CoFe_2O_4$ (red) and $Ni_{0.75}Co_{0.75}Fe_{1.5}O_4$ (blue); the data of $NiCo_2O_4$ was measured at 50 K with +40 kOe cooling field, while those of $CoFe_2O_4$ and $Ni_{0.75}Co_{0.75}Fe_{1.5}O_4$ at 20K with +70 kOe cooling field. $t_{FM} = t - t_I$ is the ferrimagnetic layer thickness with the interfacial layer thickness $t_I$ subtracted from the total thickness t. **(b)** Temperature behaviors of exchange bias $H_E$ (circles) and coercivity $H_C$ (stars) of a 10 nm $NiCo_2O_4$ (black), 10nm $CoFe_2O_4$ (red) and a 12nm $Ni_{0.75}Co_{0.75}Fe_{1.5}O_4$ (blue). **(c)** Temperature behaviors of saturation moments $M_S$ of a 10 nm $NiCo_2O_4$ (black), a 10 nm $CoFe_2O_4$ (red) and a 12nm $Ni_{0.75}Co_{0.75}Fe_{1.5}O_4$ (blue). The data about $CoFe_2O_4$ films here were extracted from [8].

## 5. Interfacial reconstruction in $Ni_xCo_yFe_{3-x-y}O_4$(111)/α-$Al_2O_3$(0001)

To reveal the interfacial reconstruction in these $Ni_xCo_yFe_{3-x-y}O_4$(111)/ $\alpha$-$Al_2O_3$(0001) thin films, we recorded thickness-resolved RHEED patterns in situ along the $Ni_xCo_yFe_{3-x-y}O_4$ $[\bar{1}10]$ in-plane direction. For each thickness, the intensity of RHEED image is summed along the streak direction; thickness-resolved RHEED pattern is then obtained by combing the results of different thicknesses, as shown in Fig. 4(a) for $Ni_{0.5}Co_{1.5}FeO_4$ as an example (see Fig. S5 for all materials in Supplementary Materials). As illustrated in Fig. 4(a), the interfacial-layer stands out without obvious RHEED lines (2 2 -4) and (-2 -2 4) of the spinel structure and the distances between (4 4 -8) and (-4 -4 8) lines in interfacial-layer are smaller than that of spinel $Ni_{0.5}Co_{1.5}FeO_4$, matching the crystal structure and lattice constants of FeO, CoO and/or NiO. To quantify the differences between the interfacial layer and the well-defined $Ni_{0.5}Co_{1.5}FeO_4$, from Fig. 4(a), we extract the thickness dependence of the in-plane lattice constant relative to the thick-limit value, and the intensity of RHEED lines (2 2 -4). Fig. 4(b) corresponds to the $Ni_{0.5}Co_{1.5}FeO_4$ film in the range 0<$t$<6.5 nm. The profiles of in-plane lattice constant and RHEED line (2 2 -4) intensity in Fig. 4(b) indicate that the interfacial layer at least includes three sublayers: one main sublayer $M_1$ (bluish region) with a larger lattice constant, a transition layer $T_1$ (greenish region) between the substrate and $M_1$, and a second transition layer $T_2$ (reddish region) between $M_1$ and the second main layer $M_2$. Both the RHEED pattern and the lattice constant of the $M_1$ sublayer match the rock-salt structure of FeO, CoO and NiO. Since PLD growth tends to keep the cation ratio from the $Ni_xCo_yFe_{3-x-y}O_4$ target, $T_1$ and $M_1$ are plausibly some mixtures of FeO, CoO and NiO according to the chemical formula $Ni_xCo_yFe_{3-x-y}O_4$. In Fig. 4(b), we defined two quantities to characterize the interfacial reconstruction: the interfacial layer thickness $\boldsymbol{t_I}$ is defined as the sum of the thicknesses of $T_1$, $M_1$ and half of $T_2$ with half of $T_2$ as the error bar, and $\Delta\boldsymbol{a_{IP}}$ as the in-plane lattice constant difference between $M_1$ and $M_2$.

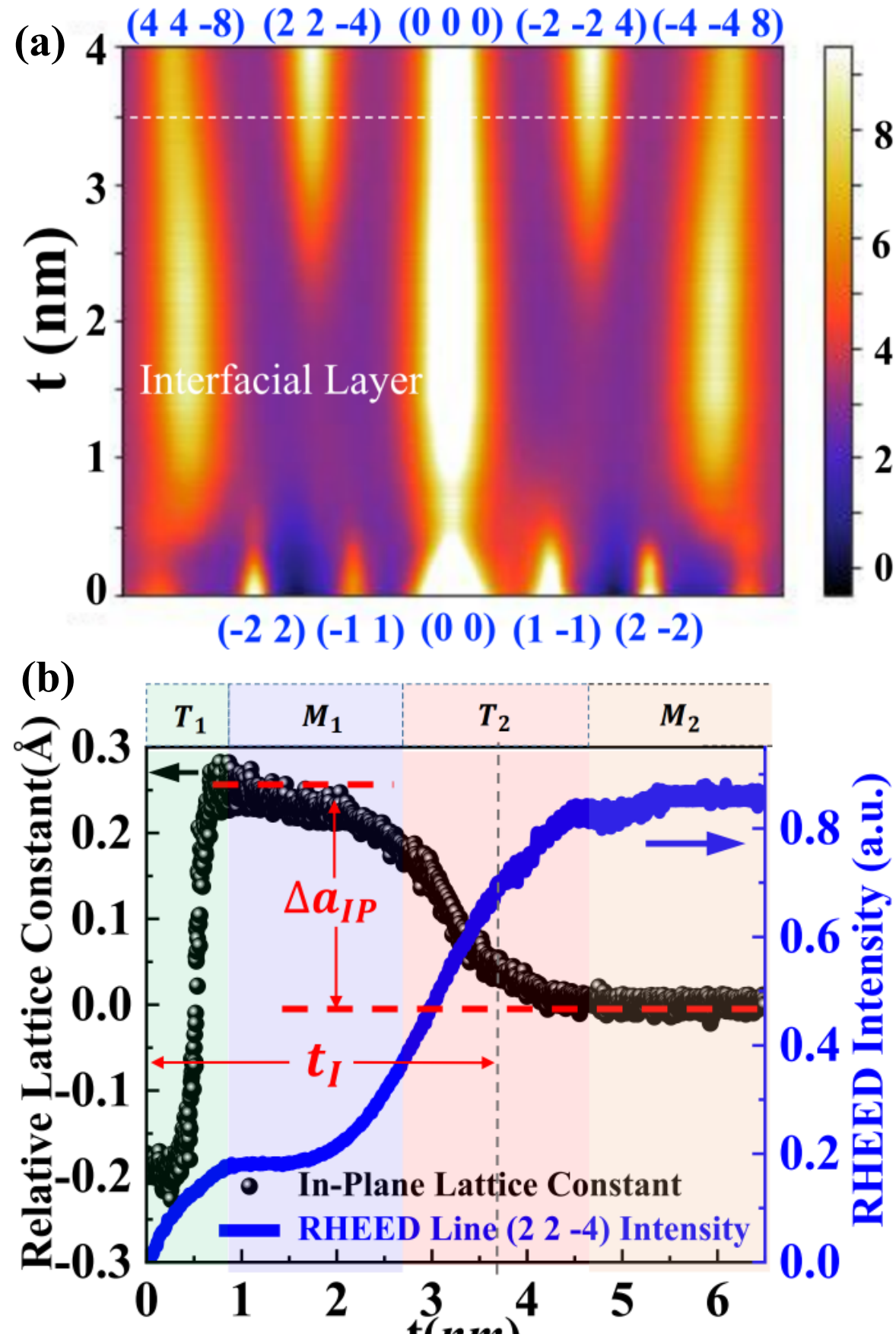

FIG. 4. **(a)** Typical thickness-resolved RHEED patterns of a $Ni_{0.5}Co_{1.5}FeO_4$ sample along the in-plane direction $[\bar{1}10]$; the white dashed lines mark the boundaries of interfacial layers defined later in (b). **(b)** Thickness-dependent relative in-plane lattice constant (black circles) and (2 2 -4) RHEED line intensity (blue curve) extracted from (a); the interfacial layer contains one main sublayer $M_1$ (bluish region) with larger lattice constant, transition layer $T_1$ (greenish region) between substrate and $M_1$ and transition layer $T_2$ (reddish region) between $M_1$ and second main layer $Ni_{0.5}Co_{1.5}FeO_4$ $M_2$; $\Delta a_{IP}$ is the in-plane lattice constant difference between the main sublayer $M_1$ and second main layer $Ni_{0.5}Co_{1.5}FeO_4$ $M_2$; The in-plane lattice constant of

## 6. The role of Co

With both magnetic and structural characterizations, we have demonstrated that an interfacial layer tends to emerge between the substrate and the ferrimagnetic $Ni_xCo_yFe_{3-x-y}O_4$ films as $0.15 \leq y \leq 2$ and contains CoO as the essential component to produce the observed exchange bias. In this section, we provide more evidence and analyses in terms of cation concentrations.

Fig. 2(a) already suggests that Co plays a key role in the mechanism of exchange bias: by comparing $NiFe_2O_4$ and $Ni_{0.95}Co_{0.15}Fe_{1.95}O_4$, slight doping of Co leads to emergence of exchange bias, while over-doping of Co eliminates the intrinsic exchange bias phenomenon again as seen from the case of $Ni_{0.15}Co_{2.55}Fe_{0.3}O_4$. To quantify the unique role of Co in the mechanism of exchange bias and interfacial reconstruction, we extract from Fig.2(a) and thickness-resolved RHEED data the dependence of $H_E$, $\boldsymbol{t_I}$ and $\Delta\boldsymbol{a_{IP}}$ on the cation concentrations and plot them in Fig. 5(a)-(c). Because the magnitudes of exchange bias values are sensitive to the specific ferromagnetic components, they vary dramatically among different $Ni_xCo_yFe_{3-x-y}O_4$ samples as we already seen in Fig.2 and Fig.S2. In Fig.5(a). We therefore focus on whether the cation concentrations correlate with the emergence of exchange bias or not, for which we indicate using P (positive) and N (null) respectively. As defined above, $\boldsymbol{t_I}$ and $\Delta\boldsymbol{a_{IP}}$ characterize structural interfacial reconstruction. Based on Fig. 5, the dependence of exchange bias and interfacial reconstruction on Ni concentration $x$ and Fe concentration $3\text{-}x\text{-}y$ exhibit no obvious trends. In particular, $\boldsymbol{t_I}$ and $\Delta\boldsymbol{a_{IP}}$ can take two considerably distinctive values in two different materials $NiFe_2O_4$ and $NiCo_2O_4$ of the same Ni concentration $x$=1. On the other hand, $H_E$, $\boldsymbol{t_I}$ and $\Delta\boldsymbol{a_{IP}}$ appear to have a consistent dependence on the Co concentration $y$: $H_E$, $\boldsymbol{t_I}$ and $\Delta\boldsymbol{a_{IP}}$ approach zero at $y$=0 and $y$=2.55, while they stay finite in the range $0.15 \leq y \leq 2$. This consistency reveals the key role of Co in the interfacial reconstruction mechanism. Moreover, since $Ni_{0.67}CoFe_{1.33}O_4$ and $CoFe_2O_4$ show similar $\boldsymbol{t_I}$ and $\Delta\boldsymbol{a_{IP}}$ values (two $y$=1 points in Fig.5(b)&5(c)), Ni and Fe concertation seem less critical in the interfacial reconstruction. The matching $\boldsymbol{t_I}$-y and $\Delta\boldsymbol{a_{IP}}$-y relations further suggest that Co concentration dominates the modulation of the lattice constant and the thickness of the interfacial layer by modulating the reconstruction process. In terms of the growth process, our results here and in [23] reveal that, when oxygen vacancy increases (low growth oxygen pressure), Co rather than Fe and Ni tends to be able to initiate the rock-salt interfacial layer over the c-plane of $\alpha$-$Al_2O_3$. The apparent counter example for $y$=2.55, might be a result of the worse crystallization in its growth as indicated by the weak and incomplete XRD (111) peaks of $Ni_{0.15}Co_{2.55}Fe_{0.3}O_4$ in Fig.1(a); it is also plausible that the reconstruction mechanism alters significantly once Co concentration surpasses certain value. Although the microscopic mechanism of the interfacial reconstruction in $Ni_xCo_yFe_{3-x-y}O_4$ thin film family requires further experimental and theoretical studies, the analysis in this section clearly supports the aforementioned exchange bias mechanism, i.e., antiferromagnetic CoO in the interfacial layer couples with ferrimagnetic $Ni_xCo_yFe_{3-x-y}O_4$ to generate the exchange bias.

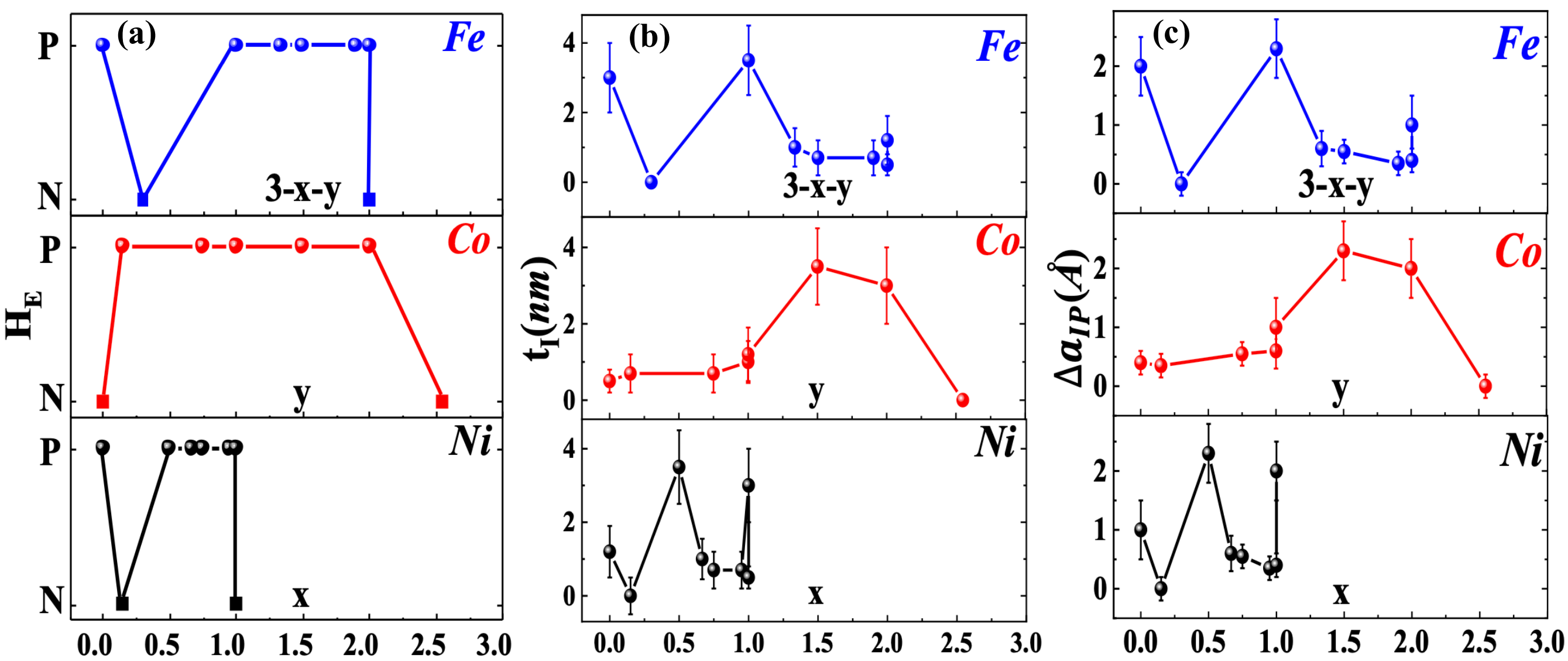

FIG. 5. The dependence of exchange bias $H_E$ (**a**), interfacial layer thickness $t_I$ **(b)** and the relative in-plane lattice constant of the interfacial layer $\Delta a_{IP}$ (c) on the concentration of Fe (top panels), Co (middle panels) and Ni (bottom panels). In (a), balls denote that substantial exchange bias larger than 0.2 kOe was observed (P) for certain thin films, while squares stand for a null exchange bias value (N) for certain thin films. $t_I$ and $\Delta a_{IP}$ were extracted from Fig. S5.

## 7. Conclusions

In conclusion, we have established a spinel oxide thin film family $Ni_xCo_yFe_{3-x-y}O_4(111)/\alpha\text{-}Al_2O_3(0001)$ to generate and study intrinsic exchange bias. The intrinsic exchange bias was demonstrated to be produced by an antiferromagnetic interfacial layer, in which the effect of CoO dominates. The exchange bias and interfacial reconstruction could be tuned greatly by many parameters including thickness, temperature, growth oxygen pressure and cation concentrations. Thanks to the versatile properties of $Ni_xCo_yFe_{3-x-y}O_4$ ranging from ferrimagnetic semimetal to antiferromagnetic insulator, this thin film family holds promising potential for applications in magnetic storage devices and spintronic devices.

## Acknowledgements

The authors acknowledge the primary support from the National Science Foundation (NSF) through EPSCoR RII Track-1: Emergent Quantum Materials and Technologies (EQUATE), Award No. OIA-2044049. The research was performed in part in the Nebraska Nanoscale Facility: National Nanotechnology Coordinated Infrastructure and the Nebraska Center for Materials and Nanoscience, which are supported by the NSF under Grant No. ECCS-2025298, and the Nebraska Research Initiative.

# Supplemental Materials for "Intrinsic Exchange Bias from Interfacial Reconstruction in an Epitaxial $Ni_xCo_yFe_{3-x-y}O_4(111)/\alpha$-$Al_2O_3(0001)$ Thin Film Family"

Detian Yang[1,2], Yaohua Liu[3] and Xiaoshan Xu[2,4,*]

[1] Shanghai Key Laboratory of High Temperature Superconductors, Department of Physics, Shanghai University, Shanghai 200444, China
[2] Department of Physics and Astronomy, University of Nebraska, Lincoln, Nebraska 68588, USA
[3] Second Target Station, Oak Ridge National Laboratory, Oak Ridge, Tennessee 37830, USA
[4] Nebraska Center for Materials and Nanoscience, University of Nebraska, Lincoln, Nebraska 68588, USA

## 1. RHEED patterns of eight materials of the $Ni_xCo_yFe_{3-x-y}O_4(111)/\alpha\text{-}Al_2O_3(0001)$ ($0\le x+y\le 3$) thin film family.

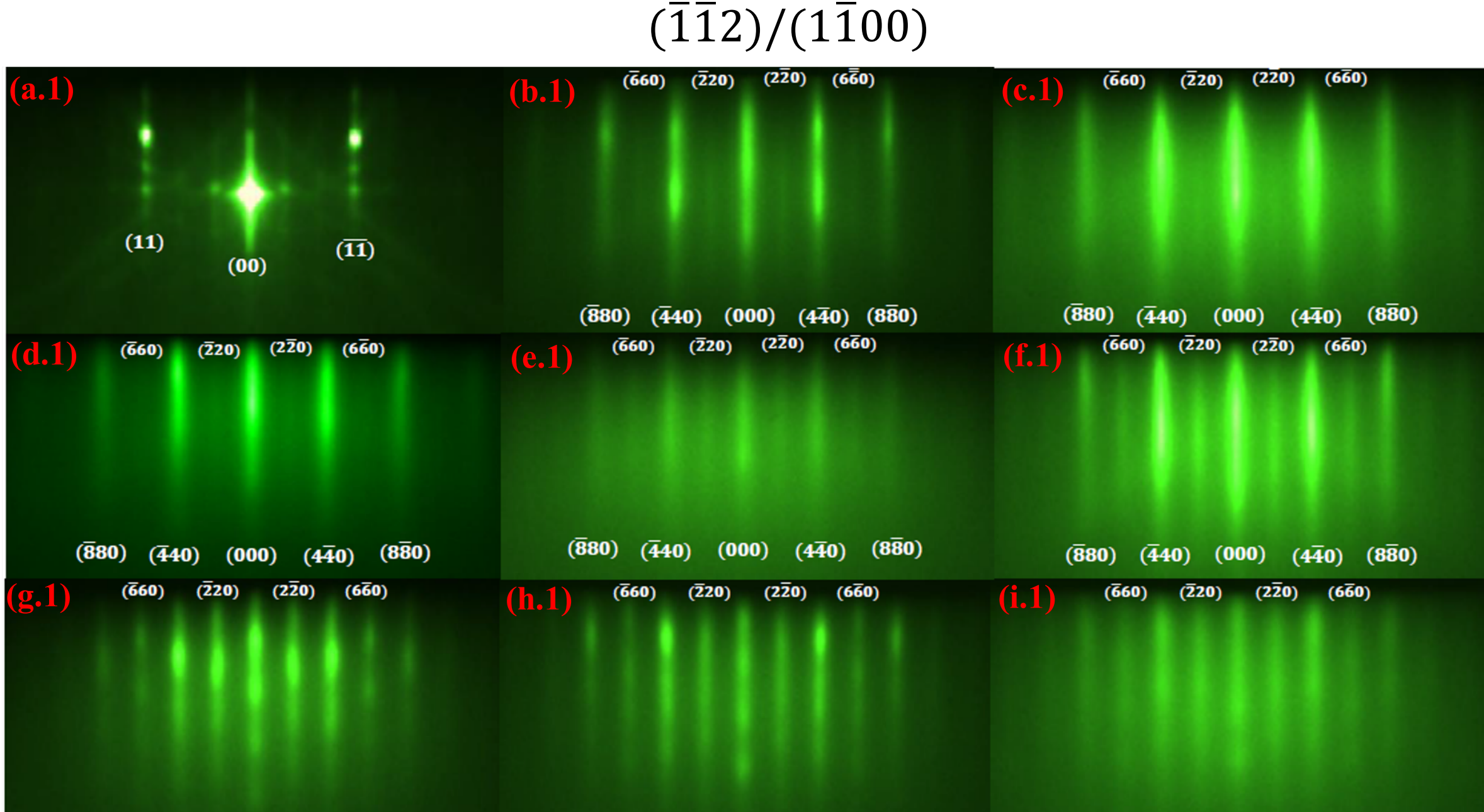

$(\bar{1}10)/(11\bar{2}0)$

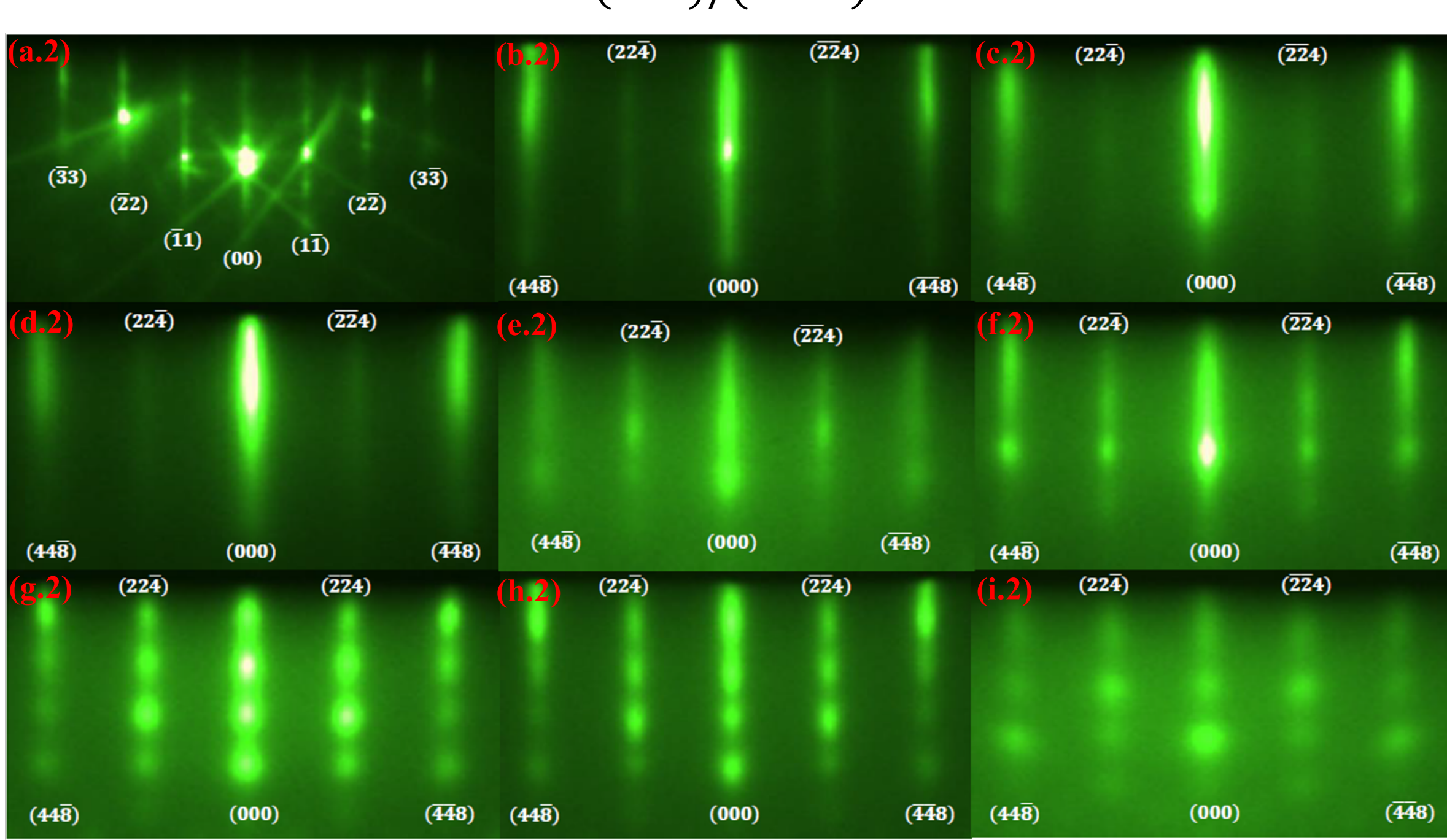

Fig.S1. Typical RHEED patterns of $Ni_xCo_yFe_{3-x-y}O_4$ thin films studied in this work along two perpendicular in-plane directions **(a-i.1)** $(\bar{1}10)/(11\bar{2}0)$ and **(a-i.2)** $(\bar{1}\bar{1}2)/(1\bar{1}00)$: **(a)** Sapphire substrate; **(b)** 11 nm $NiFe_2O_4$ ; **(c)** 18 nm $Ni_{0.95}Co_{0.15}Fe_{1.95}O_4$; **(d)** 32 nm $Ni_{0.75}Co_{0.75}Fe_{1.5}O_4$; **(e)** 21 nm $Ni_{0.67}CoFe_{1.33}O_4$ ; **(f)** 12 nm $CoFe_2O_4$; **(g)** 18 nm $Ni_{0.5}Co_{1.5}FeO_4$ ; **(h)** 15 nm $NiCo_2O_4$ ; **(i)** 18 nm $Ni_{0.15}Co_{2.55}Fe_{0.3}O_4$. All thin films were grown in 5mTorr oxygen pressure with 120±10 mJ pulsed laser energy.

**2. Typical in-plane hysteresis loops of eight materials of the $Ni_xCo_yFe_{3-x-y}O_4(111)/\alpha$-$Al_2O_3(0001)$ ($0\leq x+y\leq 3$) thin film family grown in 5mTorr oxygen pressure with 110 mJ pulsed laser energy.**

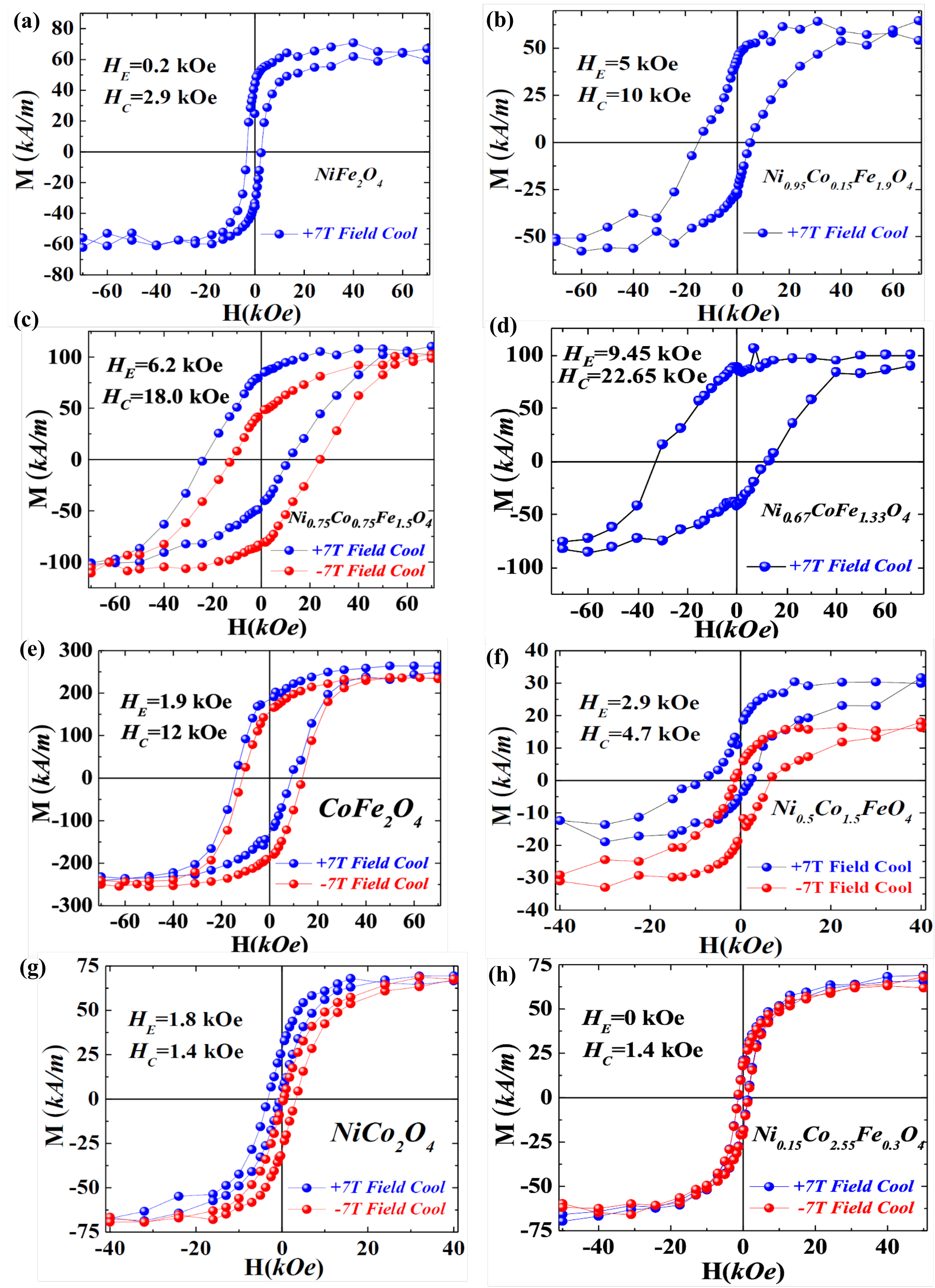

Fig.S2. In-plane hysteresis loops thin films of eight materials of $Ni_xCo_yFe_{3-x-y}O_4(111)/\alpha$-$Al_2O_3(0001)$ ($0\leq x+y\leq 3$) thin film family grown in 5mTorr oxygen pressure with 110 mJ pulsed laser energy: **(a)** 9.5nm $NiFe_2O_4$ ; **(b)** 10nm $Ni_{0.95}Co_{0.15}Fe_{1.95}O_4$; **(c)** 16nm $Ni_{0.75}Co_{0.75}Fe_{1.5}O_4$; **(d)** 21 nm $Ni_{0.67}CoFe_{1.33}O_4$ ; **(e)** 15nm $CoFe_2O_4$; **(f)** 11nm $Ni_{0.5}Co_{1.5}FeO_4$ ; **(g)** 10 nm $NiCo_2O_4$ ; **(h)** 18 nm $Ni_{0.15}Co_{2.55}Fe_{0.3}O_4$. All loops were measured at 20K after being cooled down in +7 T and / or −7 T fields. $H_E$ and $H_C$ denote exchange bias and coercivity, respectively.

## 3. Discussion on potential mechanisms of large vertical shifts of hysteresis loops in $Ni_{0.75}Co_{0.75}Fe_{1.5}O_4$ and $Ni_{0.67}CoFe_{1.33}O_4$ thin films of relatively small thicknesses

In the main text, we systematically illustrated rather a universal intrinsic exchange bias phenomenon in $Ni_xCo_yFe_{3-x-y}O_4$($0.15\leq y\leq 2$) epitaxial films which demonstrates a hysteresis loop shift normally along the field axis. However, intriguingly, we also observed large vertical shift of hysteresis loops of certain thin film members (such as $Ni_{0.75}Co_{0.75}Fe_{1.5}O_4$ and $Ni_{0.67}CoFe_{1.33}O_4$) with a small thickness of around 10 nm. According to the definition of exchange bias in the main text, the exchange bias in such samples can be as large as around10 kOe, as shown in Fig.S3(a)&(b). Apparently, the large nominal exchange bias kOe in here is a trivial result of the significant vertical shifts of hysteresis loops along the magnetization axis. Also note that, in both cases, the horizontal shift of the loops is near zero[1].

To attack such a puzzle, it is tempted to assume that the ferrimagnetic parts of the samples are not homogenous along the out-of-plane direction and contain a magnetic "hard layer" adjacent to the antiferromagnetic interfacial layer. As indicated in Fig.S3(c), the magnetic moment of the hard layer is aligned by the cooling field $H_{CF}$ and stay pinned even under an opposite external field so that the resultant hysteresis loop would be effectively vertically shifted. The pinning force can either originate from the interfacial exchange coupling from the interfacial spins of the antiferromagnetic layer or from some defects (such as antiphase boundaries) inside the hard layer. However, to support such a hard-layer model, one must resolve the discrepancy that the potential interfacial exchange coupling between the hard layer and the soft layer on the right side has not given rise to the horizontal shift of hysteresis loops in the thinner samples, but meanwhile led to considerable horizontal shift in thicker samples, as shown in Fig.S2(d) and Fig.S4.

The other plausible mechanism is a scenario proposed by M. Kiwi et al in 1999[2]. As it were, because of the smaller thicknesses of the samples, the ferrimagnetic parts could be relatively "soft" and a ferrimagnetic domain wall forms in reaction to the reversing of the external field. Due to the lower energy of Néel type domain wall[3] in thin films, as illustrated in Fig.S3(d), we assume that the domain walls emerging here are of the Néel type. Obviously, in such a case, the normal rigid-rotation assumption in the Meiklejohn-Bean model fails. And the misalignment of spins in the domain wall effectively reduces the net magnetization and would shift the hysteresis loop along the magnetization axis. Based on Kiwi's simulations and arguments[4,5], this domain-wall model could lead to a positive, negative or null exchange bias according to the magnitudes of cooling field $H_{CF}$ and the exchange coupling constant $J_{F/AF}$ between the interfacial antiferromagnetic spins and ferrimagnetic spins[5]. When $H_{CF}$ and $J_{F/A}$ takes proper values, the exchange bias can be smaller than the experimental error, and thus, the hysteresis loops turn out well-centered along the field axis. Note that, unlike the hard-layer model elaborated above, this model could explain the horizontal shift of hysteresis loops in the thicker samples. For simplicity, we assume that the uniaxial anisotropy dominates in the ferrimagnetic components of $Ni_{0.75}Co_{0.75}Fe_{1.5}O_4$ and $Ni_{0.67}CoFe_{1.33}O_4$ samples, then the domain wall width $d_w \propto \sqrt{J/K}$, where J is the exchange constant in the ferrimagnetic parts and K the uniaxial anisotropy constant. Typically, in ferri- or ferro- magnetic thin films, K would increase with the increasing of the thickness. That is, the ferrimagnetic domain wall would shrink or even disappear when it comes to thicker samples[3], restoring the rigid-rotation regime. Therefore, the exchange bias behavior of $Ni_{0.75}Co_{0.75}Fe_{1.5}O_4$ samples with multiple thicknesses still echoes the interfacial mechanism emphasized in the main text. However, in Kiwi's model, the time-reversal symmetry breaking is induced by the spin-canting of an initially compensated interfacial antiferromagnetic interfacial layer in the field cooling process. In our samples, the antiferromagnetic component was proposed as CoO and, in its bulk counterparts, it is well-established that spins mainly colinearly align parallel in the {111} lattice planes[6,7]. One plausible explanation lies in the fact that, on one hand, the magnetic order of the 1-3 nm films might substantially deviate from the bulk case; on the other, other components like FeO and/or NiO as well as defects could disrupt the magnetic order of CoO dramatically. And as a result, the interfacial surface of the antiferromagnetic layer in $Ni_{0.75}Co_{0.75}Fe_{1.5}O_4$ and $Ni_{0.67}CoFe_{1.33}O_4$ samples could be effectively compensated.

To sum up, the domain-wall model seems a more valid mechanism to have induced the giant vertical shifts observed in Fig.S3(a)&(b). Experimental techniques such as polarized neutron reflectometry could unequivocally settle the disputes here.

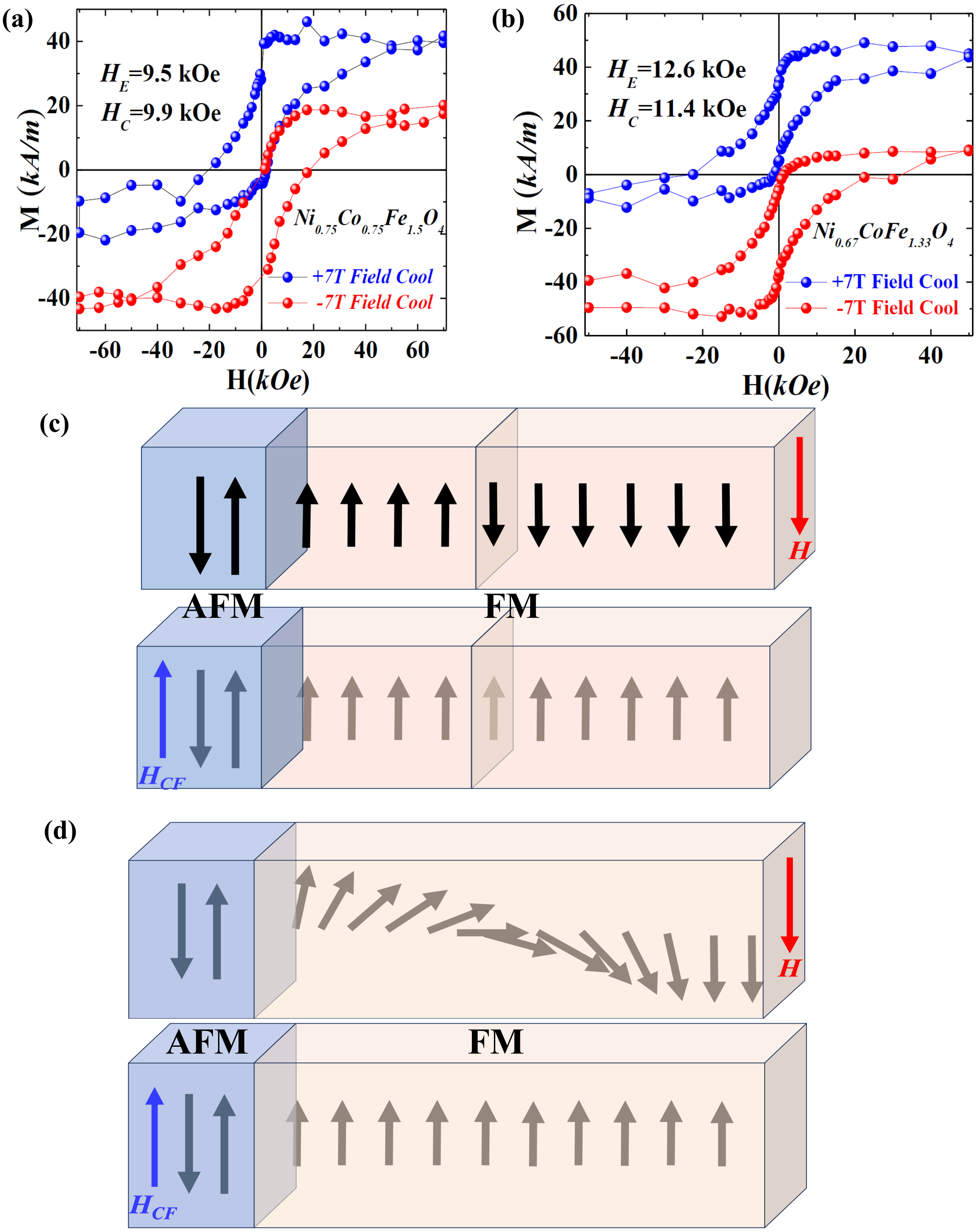

Fig.S3. In-plane hysteresis loops thin films of **(a)**10 nm $Ni_{0.75}Co_{0.75}Fe_{1.5}O_4$; **(b)** 11 nm $Ni_{0.67}CoFe_{1.33}O_4$ . Two models to explain the vertical shift of hysteresis loops in (a)&(b): **(c)** The model with a hard layer and **(d)** The ferrimagnetic domain wall model; the lower panel shows after the spin arrangement after cooling down in the field $\mathbf{H_{CF}}$ while the upper panel demonstrates the spin arrangement at the maximum external field **H** in the magnetic measurement process. All loops were measured at 20K after being cooled down in +7 T and / or −7 T fields. $H_E$ and $H_C$ denote exchange bias and coercivity, respectively. AFM denotes the antiferromagnetic interfacial layer and FM the ferrimagnetic layer.

### 4. In-plane hysteresis loops of $Ni_{0.75}Co_{0.75}Fe_{1.5}O_4(111)/\alpha\text{-}Al_2O_3(0001)$ thin films of multiple thicknesses.

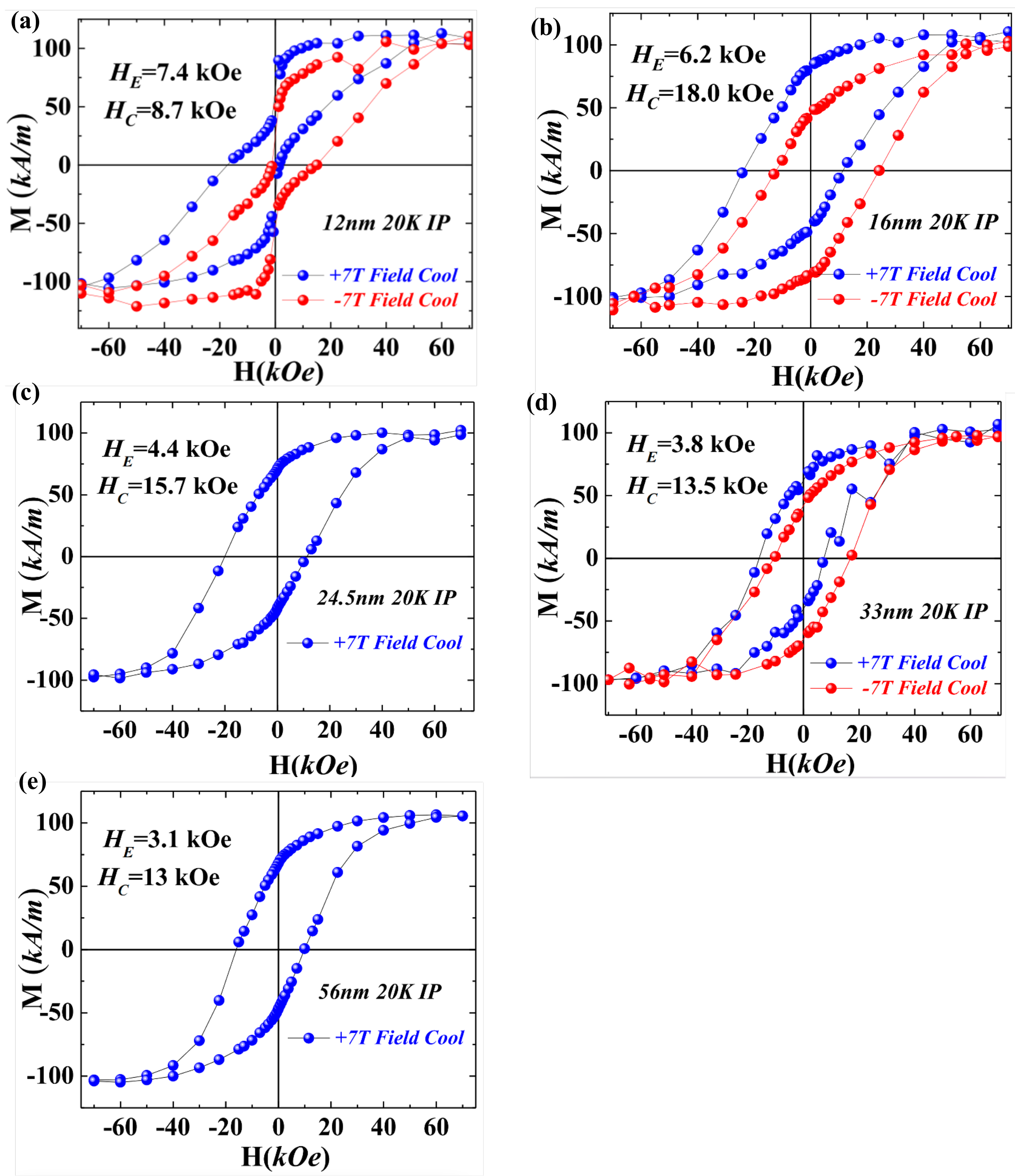

Fig.S4. In-plane hysteresis loops of $Ni_{0.75}Co_{0.75}Fe_{1.5}O_4(111)/\alpha\text{-}Al_2O_3(0001)$ thin films of multiple thicknesses and of a 21 nm $Ni_{0.67}CoFe_{1.33}O_4(111)/\alpha\text{-}Al_2O_3(0001)$ thin film: (a) 12nm; (b) 16nm; (c) 24.5nm; (d) 33nm ; (e) 56nm;(f) 21nm $Ni_{0.67}CoFe_{1.33}O_4$.All loops were measured at 20K after being cooled down in +7 T and / or −7 T fields. $H_E$ and $H_C$ denote exchange bias and coercivity, respectively. IP stands for In-plane.

## 5. Typical thickness-resolved RHEED patterns of eight materials of the $Ni_xCo_yFe_{3-x-y}O_4(111)/\alpha\text{-}Al_2O_3(0001)$ ($0\le x+y\le 3$) thin film family.

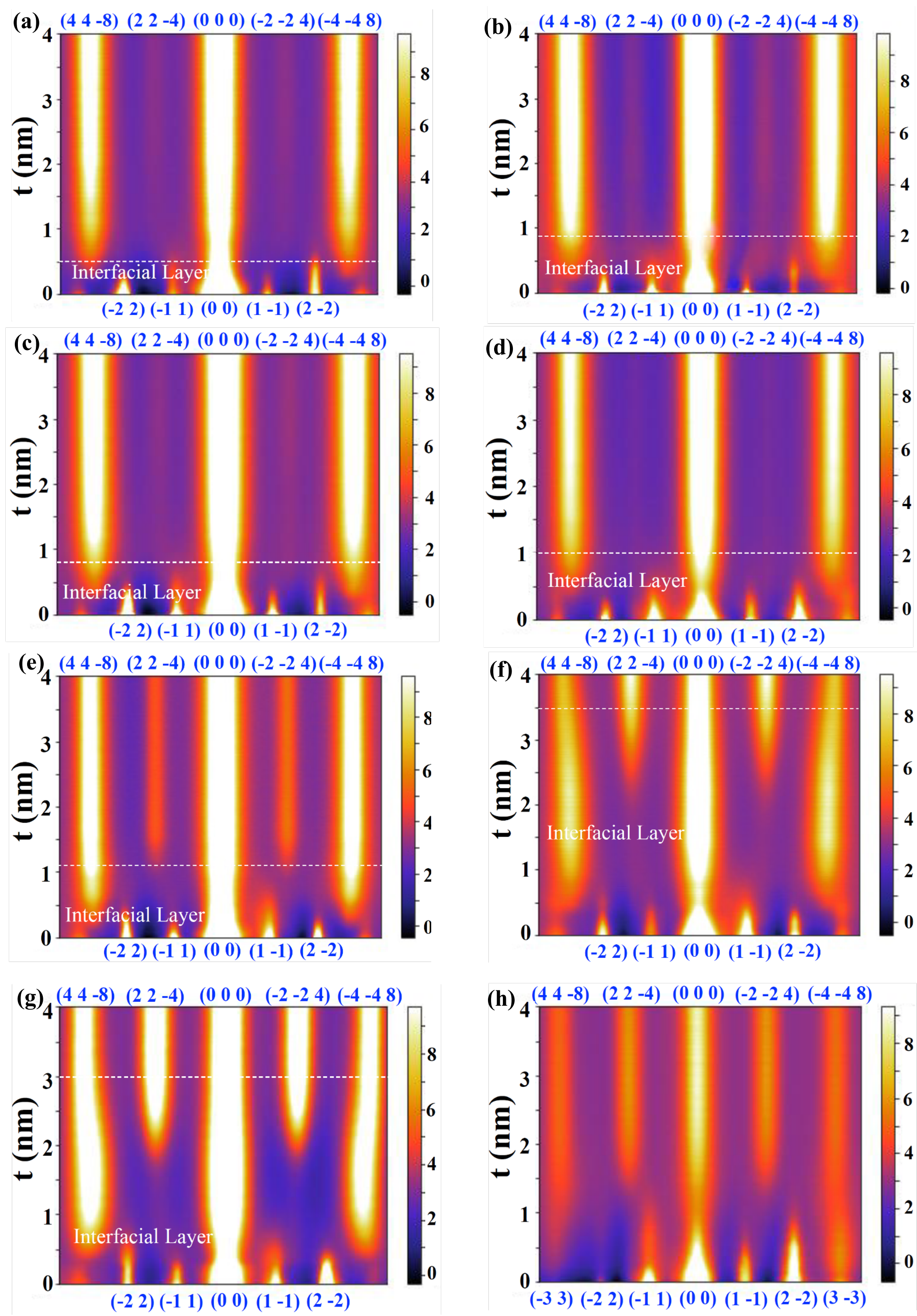

Fig.S5. Typical thickness-resolved RHEED patterns ($0\le t \le 4\ nm$) of eight materials $Ni_xCo_yFe_{3-x-y}O_4(111)/\alpha\text{-}Al_2O_3(0001)$ ($0\le x+y\le 3$) thin film family along the in-plane direction [$\bar{1}10$]. **(a)** $NiFe_2O_4$ ; **(b)** $Ni_{0.95}Co_{0.15}Fe_{1.95}O_4$; **(c)** $Ni_{0.75}Co_{0.75}Fe_{1.5}O_4$; **(d)** $Ni_{0.67}CoFe_{1.33}O_4$ ; **(e)** $CoFe_2O_4$; **(f)** $Ni_{0.5}Co_{1.5}FeO_4$ ; **(g)** $NiCo_2O_4$ ; **(h)** $Ni_{0.15}Co_{2.55}Fe_{0.3}O_4$. The white dashed lines mark the boundaries of interfacial layers defined as in Figure 4(b) in the main text. All eight materials were grown in the 5mTorr oxygen pressure with the 110 mJ pulsed laser energy.